# Determination of internal density profiles of smart acrylamide-based microgels by SANS: A multi-shell reverse Monte-Carlo approach

Marian Cors[1, 2], Lars Wiehemeier[1], Yvonne Hertle[1], Artem Feoktystov[3], Fabrice Cousin[4], Thomas Hellweg[1*], Julian Oberdisse[2*]

[1] Department of Physical and Biophysical Chemistry, Bielefeld University, Universitätsstr. 25, 33615 Bielefeld, Germany
[2] Laboratoire Charles Coulomb (L2C), University of Montpellier, CNRS, 34095 Montpellier, France.
[3] Forschungszentrum Jülich GmbH, Jülich Centre for Neutron Science JCNS at Heinz Maier-Leibnitz Zentrum MLZ, 85748 Garching, Germany.
[4] Laboratoire Léon Brillouin, UMR 12 CEA/CNRS, CEA Saclay, 91191 Gif Sur Yvette, France

* Authors for correspondence : thomas.hellweg@uni-bielefeld.de, julian.oberdisse@umontpellier.fr

## Abstract

The internal structure of nanometric microgels in water has been studied as a function of temperature, crosslinker content, and level of deuteration. Small-angle neutron scattering from poly($N$-isopropylmethacrylamide) (pNIPMAM, volume phase transition $\approx$ 44 °C) microgel particles of radius well below 100 nm in $D_2O$ has been measured. The intensities have been analyzed with a combination of polymer chain scattering and form-free radial monomer volume fraction profiles defined over spherical shells, taking polydispersity in size of the particles determined by AFM into account. A reverse Monte Carlo optimization using a limited number of parameters was developed to obtain smoothly decaying profiles in agreement with the experimentally scattered intensities. Result are compared to the swelling curve of microgel particles in the temperature range from 15 to 55 °C obtained from photon correlation spectroscopy (PCS). In addition to hydrodynamic radii measured by PCS, our analysis provides direct information about internal water content and gradients, the strongly varying steepness of the density profile at the particle-water interface, the total spatial extension of the particles, and the visibility of chains. The model has also been applied to a variation of the crosslinker content, $N,N'$-methylenebisacrylamide (BIS), from 5 to 15 mol%, providing insight in the impact of chain architecture and crosslinking on water uptake and on the definition of the polymer-water interface. The model can easily be generalized to arbitrary monomer contents and types, in particular mixtures of hydrogenated and deuterated species, paving the way to detailed studies of monomer distributions inside more complex microgels, in particular core-shell particles.

## Introduction

Smart microgels are stimuli-responsive colloids with an internal gel structure. They can be made of cross-linked poly-acrylamides exhibiting a lower critical solution temperature (LCST)[1-7] also called volume phase transition temperature, awarding them the capability to react on temperature changes. Poly($N$-isopropylacrylamide) (pNIPAM) based microgels are certainly the best studied systems at the moment, but also poly($N$-isopropylmethacrylamide) (pNIPMAM) and poly($N$-n-propylacrylamide) (pNNPAM) are of growing interest.[8-12] A combination of these monomers can be used to produce particles with variable LCST or with more complex architectures like e.g. core-shell.[9,13] Microgel particles are very promising with respect to applications in biomedical devices[14,15], as sensors[16,17], for drug delivery[18-23], as functionalized and possibly double responsive microgels for nanoreactors[24-27], and in colloidal crystals[28-34]. The possibility of creating complex microgel architectures is accompanied by the need for structural characterization. This may be achieved by small-angle neutron scattering (SANS), which is particularly useful in suspension, as a function of temperature, allowing the simultaneous determination of both average particle sizes and polydispersity, and local chain structure. Microgel structure has soon been recognized to be spatially inhomogeneous due to differences in crosslinker and monomer consumption,[35] introducing a gradual radial decrease of density and a fuzziness of the microgel surface, unless special care is taken to distribute the crosslinker more evenly[36]. The first of the fuzzy sphere models for small-angle scattering was based on a description of spherical block copolymer micelles by Pedersen and Svaneborg[37,38]. It was applied by Stieger et al. to PNIPAM microgel particles.[39] Note that in this model and all following ones, a Lorentzian term was added to the form factor describing the particle density in order to account for the polymer mesh and extract its correlation length $\xi$. The fuzzy particle density profile is obtained by convoluting the profile of a homogeneous sphere with a Gaussian of width $\sigma$, which in reciprocal space introduced simply a multiplication by a Gaussian, thus weakening the oscillations of the form factor. A more versatile alternative model, a piecewise parabolic profile, was proposed by Berndt et al.[40,41], and is referred to as 'fuzzy-sphere' model here unless otherwise mentioned. It represents the radial profile as a succession of constant densities, and parabolic decays or increases, and can thus be generalized to multi-shell structures ("fuzzy core-shell"), as proposed by Berndt et al. for pNIPMAM-core and PNIPAM-shell microgels, both hydrogenated.[40,41] The same group later investigated the effect of shell thickness and crosslinking, and analyzed the resulting density profiles in terms of expansion or compression induced by the shell.[9] One may note that non spherically-symmetric microgels have been synthesized by copolymerization. The resulting nanophase-separated structures at intermediate temperatures made of the collapsed component were described using a 'dirty snowball' model[10]. The influence of synthesis conditions, in particular surfactant ratio to tune particle size, on the morphology of PNIPAM microgels have also been investigated by using a combination of static light scattering (SLS) and laboratory small-angle X-ray scattering (SAXS) analyzed by a Gaussian fuzzy-sphere model by Deen et al.[42] Similarly, the effect of pressure jumps has been described recently also with the Gaussian fuzzy-sphere model.[43] For comparison, the final particle size of pNIPMAM microgels has been systematically varied via the surfactant and analyzed by direct imaging.[44] Coming back to scattering, a fuzzy-sphere model has been applied to hollow microgel particles obtained by removing an inorganic (usually silica) precursor nanoparticle from the center of the microgel.[45] In presence of the silica core, contrast variation using $H_2O/D_2O$ has been employed with SANS to differentiate the inorganic core from the shell. A further step was taken by adding another polymer shell, forming core-shell-shell or, after removal of the core, hollow-shell-shell particles.[46] Resulting density profiles of such hollow multi-shell microgel particles obtained by analysis of SANS data have been shown to compare favorably to MD simulations.[47] Moreover, the function of the two shells could be identified, the inner shell setting the permeability allowing control of encapsulation and release, whereas the outer one provided colloidal stability in the temperature range of interest. Similarly, the model was used to examine shell interpenetration, and in particular the effect of shell inversion, by Brugnoni et al.[48] The incorporation of guest molecules in an- and cat-ionically modified core-shell microgel particles was also followed by SANS (among other techniques), using the fuzzy core-shell model.[49] Very recently, Keidel et al. have also applied piece-wise parabolic profiles to time-resolved SAXS experiments (in parallel with computer simulations showing non spherical features), where a change in solvent provoked the transition from a swollen gel to a globule, passing through a hollowish shell intermediate structure.[50]

A different model based on a Flory-Rehner approach was recently proposed by Boon and Schurtenberger, allowing them to investigate non-homogeneous crosslinker distributions in PNIPAM microgels.[51] Together with the Gaussian fuzzy-sphere model, it was used to describe density distributions obtained by super resolution microscopy.[52] In these experiments, the fluorophore was freely diffusing inside the polymer network giving rise to localization densities probing the gel network density. For higher crosslinker contents the fuzzy-sphere model had to be modified slightly to account for a linear density increase observed in the dSTORM experiments, but then performed better than, e.g. the Boon-Schurtenberger approach. In a different study, the fluorophore was added to a core-shell microgel either using ester chemistry on primary amines, or Michael addition on shell thiols, and spatial distributions could be obtained by direct imaging, by Gelissen et al.[53] Very recently, the Siemes et al have used a photoswitchable crosslinker in a PNIPAM microgel network in order to visualize its density by super-resolution fluorescence imaging without modifying the network.[54] Remarkably, the resulting crosslinker density profile has a more smoothly deyaying shape than the monomer profile obtained by the fuzzy-sphere model.

The different versions of fuzzy-sphere models developed in the references mentioned above have in common that a fixed shape-function with e.g. a predefined choice of the number of shells is used as a starting point, with some latitude given by the thickness and position of the interfaces. In some situations, unexpected results may necessitate a modification of the model, by adding an additional layer. For instance, in the case of the silica-acrylamide particles discussed above, good fits have been obtained only by adding a dense polymer layer near the nanoparticle surface.[45] Thus, any a priori unexpected density profile is difficult to describe in general by such predefined profiles.

A form-free analysis of small-angle intensities in general has been proposed in the past, and the indirect Fourier transform by Glatter et al. can be looked at as the paradigm of such approaches.[55,56] It amounts to finding a correlation function the scattering of which coincides with the measured one. There are numerous ways to find such solutions, including random search schemes running under the name of reverse Monte Carlo (RMC).[57,58] These methods are usually applied to liquid state measurements, but have been extended to soft condensed matter, namely colloids[59], or micelles[60]. Recently, Hansen has revisited the estimation of density distributions from small-angle scattering, analyzing the influence of remaining background and different regularizations, and applied it to typical soft aggregates, like micelles and proteins.[61] In some cases, the density distributions can be defined by a small set of parameters, like cross-section density profiles in lamellar systems,[62] or radial density profiles in globular systems.[63] The densities are described as a set of Gaussian functions, and the technique is termed Gaussian deconvolution. We have applied a similar approach to monodisperse block copolymer micelles in the past,[64] but to the best of our knowledge this form-free technique has not been generally applied to small-angle scattering of polydisperse microgel particles to retrieve the radial density profile, with the exception of the work by Keidel mentioned above. Finally, it may be worth mentioning a Matrioschka-like model which was used to analyze AFM force measurements.[65] This model has some similarities with the approach in the present contribution.

Small-angle scattering is thus well-suited to study microgel morphology. This is particularly true for the possibility to highlight selected regions of microgel particles using SANS and deuteration. An example has been discussed above,[45] where the introduction of different scattering length densities via an inorganic core facilitates the identification of contributions, namely of the unexpected dense surface layer. SANS with isotopic substitution is also a method of choice for the study of individual microgel particles in a dense assembly. Scotti et al. have used this property to study the radial density profile during the squeezing of microgels in overcrowded environments, in spite of deviations from sphericity.[66] In general, SANS and deuteration can be used to unravel the spatial distribution of monomers, as for instance in core-shell particles with one component isotopically labelled, which is the ultimate aim of the approach proposed in this article, using a form-free approach to retrieve radial density profiles.

In this article, we investigate the density profiles of a system based on the monomer *N*-isopropyl methacrylamide (NIPMAM) which exhibits an LCST of ca. 44 °C[41,67,68], and which will be used as the core of a deuterated core-shell particle in the future. The outline of the paper is to first present results from PCS and direct imaging providing additional information on shape and polydispersity needed to fix parameters for the following SANS analysis. The form-free SANS-model, which is analogous to layer models used in reflectometry, is then presented, and the resulting profiles discussed for a temperature- and a crosslinker-series, as well as deuteration, and compared to existing models.

## Experimental Section

**Synthesis:** The synthesis and purification of the microgel systems studied in this article was described in detail elsewhere.[69,70] In short, we did a precipitation polymerization under nitrogen atmosphere to synthesize microgels with different crosslinker contents (CC). All chemicals were purchased from Sigma-Aldrich. The monomer was *N*-isopropylmethacrylamide (NIPMAM, purity >97%), the cross-linker *N*,*N'*-methylenebisacrylamide (BIS, 99%), the initiator ammonium peroxodisulfate (APS, >98%), and the ionic surfactant sodium dodecyl sulfate (SDS, >99%). The amount of monomer was (12.4 mmol, 1.56251 ± 0.00054 g). With respect of the amount of monomer the amount of initiator (3.3 mol%, 0.09252 ± 0.00010 g), and surfactant (6 mol%, 0.21260 ± 0.00027 g) were constant. The amount of crosslinker with respect to the amount of monomer was 0.09467 ± 0.00010 g (5 mol%, 0.61 mmol), 0.18943 ± 0.00010 g (10 mol%, 1.23 mmol), 0.28435 ± 0.00010 g (15 mol%, 1.84 mmol). Thus we have three samples of pNIPMAM microgels with different CC of 5, 10 and 15 mol%. The microgel with a CC of 10 mol% was also synthesized partially deuterated with D7-pNIPMAM (Polymer Source. Inc., Quebec, Canada) and fully deuterated with D12-pNIPMAM (Polymer Source. Inc., Quebec, Canada). The purification was done by centrifugation. The expected volume phase transition temperature of microgel particles made of this monomer in water is 44 °C.[41,67,68]

**Dynamic light scattering:** Temperature-dependent photon correlation spectroscopy (PCS) reported in this article was performed at a fixed angle of 60°, as microgel motion at low concentrations is diffusive and unperturbed by internal network modes (see SI for an exemplary angular scan from 30° to 120°). The light source of the partly homemade set-up was a He-Ne laser (HNL210L-EC, 632.8 nm, Thorlabs, Newton, USA), a multiple tau digital correlator (ALV-6010, ALV GmbH, Langen, Germany), and a detector (SO-SIPD, ALV GmbH, Langen, Germany). The sample was placed in a decalin matching bath. For temperature control a refrigerated bath (Haake C25P, Thermo Fisher Scientific, Waltham, USA) with a controller (Phoenix II, Thermo Fisher Scientific, Waltham, USA) was used.

**Transmission Electron Microscopy (TEM):** TEM images were taken with a JEM-2200 (JEOL, Freising, Germany) equipped with a cold field emission electron gun and an omega filter which was used to obtain more contrast. The acceleration voltage was 200 kV. The sample preparation was done at room temperature. A droplet (3 µL) of highly diluted (ca 0.0001 wt%) sample in water was dropped on a carbon coated copper grid (ECF200-Cu, 200 mesh, Science Services, Munich, Germany) and after 1 min. the water excess was blotted off with a filter paper, and samples dried in air over a day.

**Atomic force microscopy (AFM):** The AFM measurements (Bruker, Billerica, USA) were done in the tapping mode using a Controller (NanoScope V, Veeco, Plainview, USA) and a silicon tip (RTESP, Bruker, Billerica, USA) at a frequency of 318 kHz. For the sample preparation a droplet (3 µL) of highly diluted (ca 0.0001 wt%) microgels in water was dropped on a silicon wafer and dried overnight. The AFM measurements allowed the determination of the polydispersity in size of the dried microgels, needed as input for the Monte-Carlo simulations. The polydispersity is described with a Gaussian distribution function of the number of microgels per unit volume of a radius comprised between r and r+dr.

$$P(r, R, \sigma)dr = \frac{1}{\sigma\sqrt{2\pi}} e^{\frac{(r-R)^2}{2\sigma^2}} dr \tag{1}$$

Throughout this article, we will refer to the relative polydispersity σ/R.

**Small-angle neutron measurements:** The SANS measurements were done at KWS-1 (JCNS at MLZ, Garching, Germany)[71] and at PA20 (LLB, Saclay, France)[72]. Three configurations were used on KWS-1: 1 m and λ=5 Å, 8 m and 5 Å, and 20 m and 12 Å. The configurations at PA20 were: 1.5m and 4 Å, 8 m and 6 Å, and 20 m and 6 Å. The overall q-range was from 0.001 Å$^{-1}$ to 0.3 Å$^{-1}$. The data from MLZ were normalized (detector electronic noise, water or Plexiglas, empty cell) using QtiKWS (JCNS, Germany), and those of LLB by Pasinet software, and data reduction was performed with OriginLab.

The scattering length densities (SLD) of all ingredients have been estimated assuming a density 1.107 g/cm$^3$ for D$_2$O and of 1.0 g/cm$^3$ for all other ingredients. For the monomers (repeat units), crosslinker, and solvents, the following values were obtained: $\rho_{mono}(C_7H_{12}NO) = 0.80\ 10^{10}$ cm$^{-2}$, $\rho_{CC}(C_7H_8N_2O_2) = 1.86\ 10^{10}$ cm$^{-2}$, $\rho_{H2O} = -0.56\ 10^{10}$ cm$^{-2}$, and $\rho_{D2O} = 6.38\ 10^{10}$ cm$^{-2}$. Given the low amount of the crosslinker and the close ρ-values of the hydrogenated constituents of the microgel, they are difficult to distinguish by scattering, and we have combined them in a single fictitious monomer of weighted average mass and scattering length density, called 'monomer' for simplicity throughout this article. Then a contrast variation experiment has been performed on the microgel containing 10 mol% CC, varying the solvent from $H_2O$ to $D_2O$. The sign-corrected square-root of the coherent intensities for different q-values is reported in Figure 1. The match-point is found to be $1.0 \cdot 10^{10}$ $cm^{-2}$ corresponding to the SLD of the fictitious monomer to be used in this simulation. It corresponds to a density of the fictitious monomer of 1.1 $g/cm^3$, close to our above estimate, and agrees with other studies of microgel densities.[39] Note that the simulation described below is designed to accommodate also monomers of very different scattering length density, e.g. deuterated ones, describing them independently. A preliminary experiment has been performed with deuterated monomers D7-pNIPMAM and D12-pNIPMAM. Their scattering length densities have been determined with a contrast variation similar to the one shown in Figure 1.

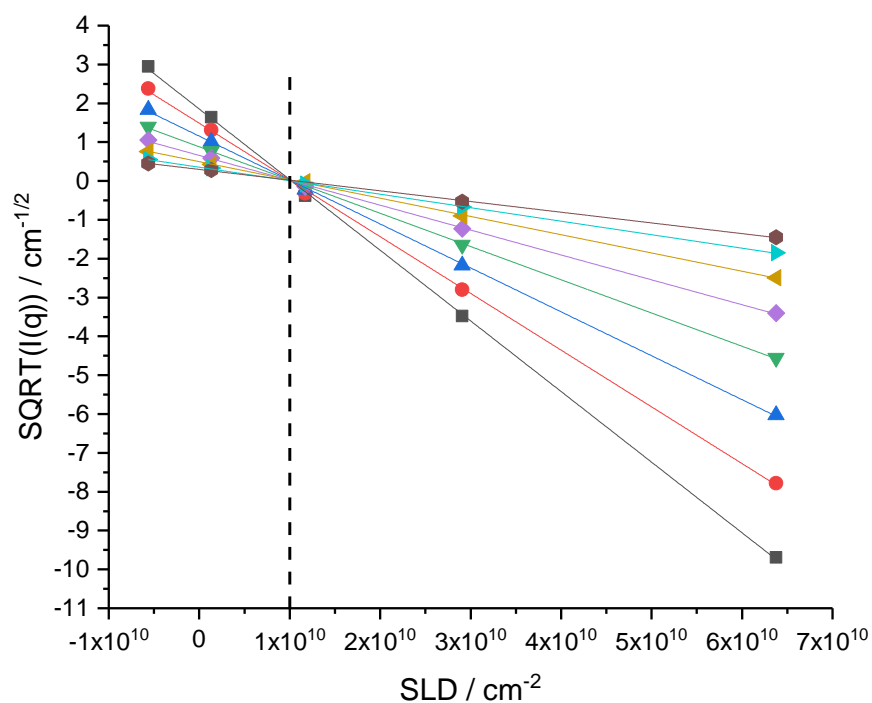

**Figure 1:** Contrast variation of pNIPMAM microgels (10 mol% CC, c=1 wt%) in water. The sign-corrected square-root of the coherent intensity is given as a function of solvent scattering length density varied from $H_2O$ to $D_2O$ (-0.561, 0.1326, 1.173, 2.907, 6.375)·$10^{10}$ $cm^{-2}$, for eight q-values (see SI). The match-point is indicated by the dotted line ($\rho_{microgel} = 1.0 \cdot 10^{10}$ $cm^{-2}$).

**Analysis of SANS data.** The comprehensive description of the small-angle scattering intensity of non-interacting microgels in terms of a general monomer volume fraction profile of the particles $\Phi(r)$ as a function of their radius is the general goal of this article. There are three contributions to the form-free model developed here, and one main assumption. The latter is that it is supposed that the microgels are spherically symmetric, allowing to define the density on spherical shells – evidence for symmetry will be shown in the results section. The first contribution to the model is an ad-hoc addition of chain scattering. Chain scattering appears in the intermediate- to high-q range, and it will be shown that the details of its description do not influence the main conclusions of this article, due to the separation of length scales. Moreover, the associated parameters, namely the chain radius of gyration $R_g$, the high-q exponent $\nu$ related to the chain statistics, and the general prefactor $I_0^{chain}$, will be shown to be correlated with the experimental conditions, namely good/poor solvent properties. The chain contribution $I_{chain}(q)$ reads:

$$I_{chain}(q) = I_0^{chain}\left[\left(\frac{1}{\nu U^{\frac{1}{2\nu}}}\right)\gamma\left(\frac{1}{2\nu}, U\right) - \left(\frac{1}{\nu U^{\frac{1}{\nu}}}\right)\gamma\left(\frac{1}{\nu}, U\right)\right] \tag{2a}$$

with

$$U = (2\nu + 1)\cdot(2\nu + 2)\cdot\frac{(qR_g^C)^2}{6} \tag{2b}$$

$$\gamma(a,b) = \int_0^b t^{a-1}e^{-t}dt \tag{2c}$$

where we have used the function describing the generalized coil[73]. For $\nu$ = ½, Eq. (2a) reduces to the standard Gaussian coil described by a Debye function, which corresponds to the well-known fractal dimension of the chain d = $1/\nu$ = 2.[74,75] One may note that the chain scattering corresponds to connecting subchains forming the entire microgel. Alternatively, one could have described the chain scattering using an Ornstein-Zernike approach, potentially modifying empirically the power law in order to fit the high-q decrease, whereas Eqs. (2) is based on the true statistics[73], which is why it is preferred. Then the size parameter $\xi$ would give the meshsize, in analogy with $R_g$ used in Eqs. (2).

The next and main ingredient of this model is the monomer density profile $\Phi(r)$ of a monodisperse microgel, which is adapted automatically to the scattered intensity as described in the results section. As SANS is a low resolution method, this profile is defined only in a coarse-grained manner, choosing $N_p$=30 homogeneous shells of fixed thickness set to $\Delta R$ = 2 nm as building units. The profile $\Phi(r)$ can be transformed into the scattering length density $\rho(r)$, using the monomer scattering length densities given in Table 1:

$$\rho(r) = \rho_{Mono}\cdot\Phi(r) + \rho_{Solv}\cdot\left(1 - \Phi(r)\right) \tag{3a}$$

This confers contrast to the microgel in the solvent, $\Delta\rho(r) = \rho(r) - \rho_{Solv}$, and gives rise to the scattered intensity I:

$$I(q,r) = \frac{N}{V}\left[\int 4\pi r^2\Delta\rho(r)\frac{sin(qr)}{qr}dr\right]^2 \tag{3b}$$

Here N/V is the number of microgels per volume chosen to satisfy the concentration of monomer volume. In our reverse Monte-Carlo (RMC) implementation, we translate Eq. (3b) into a discrete sum over the shells, with the radial variable r given by the number of the shell i, thus the outer radius of each shell r(i) = i$\Delta$R:

$$I_{RMC}(q,r,n) = \sum_{i=1}^{N_p} \frac{N}{V}\left[\frac{4\pi}{3}r(i)^3\big(\Delta\rho(i) - \Delta\rho(i+1)\big)3\frac{sin(qr) - qrcos(qr)}{(qr)^3}\right]^2 \qquad (3c)$$

Here $\Delta\rho$(i) is the scattering length contrast of shell i, and the contrast of the surrounding solvent shell $\Delta\rho$(N$_p$+1) is set to zero.

The last contribution is microgel size polydispersity. Our description accounts for both the wavelength smearing of FWHM 10% in SANS (which dominates the resolution function in the q-range of interest) and the microgel polydispersity in size measured by AFM (Gaussian $\sigma$ = 20.0 %) in a single parameter $\sigma$, in practice $\sigma$ = 20.4 % (see SI for details). In absence of interactions, the form factor can be averaged over the different sizes, using a polydispersity integral over the number distribution function P(r, R, $\sigma$) following Eq. (1), which in practice is reduced to a finite sum over seven sizes embracing the total size distribution (see Figure S1 for illustration). The total intensity then reads:

$$I = \int I_{RMC}(q,r)P(r,R,\sigma)dr + I_{chain}(q) \qquad (4)$$

Note that the polydispersity is introduced by stretching or compressing the density profile, via a modification of the shell thickness $\Delta$R, while keeping the density identical. Bigger microgels have thus thicker shells, higher masses, and contribute more to the scattering following the weighting in Eq. (4).

The total list of parameters of this model reads: (a) the three chain parameters, out of which $\nu$ is determined directly from the slope of the high-q part of the scattering curve, and I$_0$$^{chain}$ and R$_g$ which are coupled, and the exact values of which do not influence the results of the model (see Figure S6 for maximum and minimum R$_g$ values; also Figure S7). (b) the polydispersity which is fixed by AFM. (c) The total number of monomers in a microgel, which is fixed from the low-q intensity and concentration. (d) The parameters of the profile, namely the thickness of each shell, and the total size. While the latter is chosen big enough such that there are always empty shells outside the particles, the exact value of the thickness does not influence the result. Shells were chosen to be thick enough to keep information content adequate, and thin enough to describe correctly the profile intensity. In practice, the maximum resolution of our experiments of typically 2$\pi$/q$_{max}$ gives a useful order of magnitude of the parameter $\Delta$R, set to 2 nm. (e) Finally, the profile itself corresponds to some twenty values, which are fixed by minimizing the deviation between fit and measured intensity under the constraint of a smoothly decaying profile. Note that the intensity contains a comparable amount of information.



As a last comment, it should be noted that sometimes interactions between microgels cause a very weak structure factor visible in the low-q range, in particular at low temperature where microgel particles are swollen. Therefore, samples have been prepared at very low volume fractions (ca. 0.08%). We have checked that the structure factor does not affect our analysis due to the possibility of describing the same particles at high temperature first, where particles are small and do not interact at such concentrations.

**Results and Discussion**

**Sphericity, polydispersity, and swelling curve.** Before proceeding with the SANS analysis of the internal structure of the microgels described by the monomer profile, as a function of crosslinker content and temperature, the assumption of an isotropic profile has to be checked. This assumption makes sense only if the microgels are indeed spheres. In Figure 2a, a representative real-space picture of the microgels with 10 mol% crosslinker content (CC) obtained by TEM is presented. This is a 2D projection of the particles in the dried state, and due to low contrast with the carbon grid, the interfaces are not well resolved. In Figure 2b, an AFM result is presented, with sharp enough contrast to determine the distribution of particle surfaces, and thus sizes (see SI). The microgels have the typical pancake-like structure (diameter 84 nm, height: 5 nm). This suggests that the particles in suspension exhibit spherical shape, because anisotropic shapes would result in anisotropic projections. Moreover, we have shown that particles at interfaces and in bulk have a similar swelling behavior,[69] suggesting that internal heterogeneities do not affect measurements on surfaces. In AFM, the particles have a mean radius of 42 nm, and a polydispersity of $\sigma$ = 20.0 %. From the pictures in Figure 2 it appears to be necessary to perform SANS measurements in order to resolve the internal structure of the microgels.

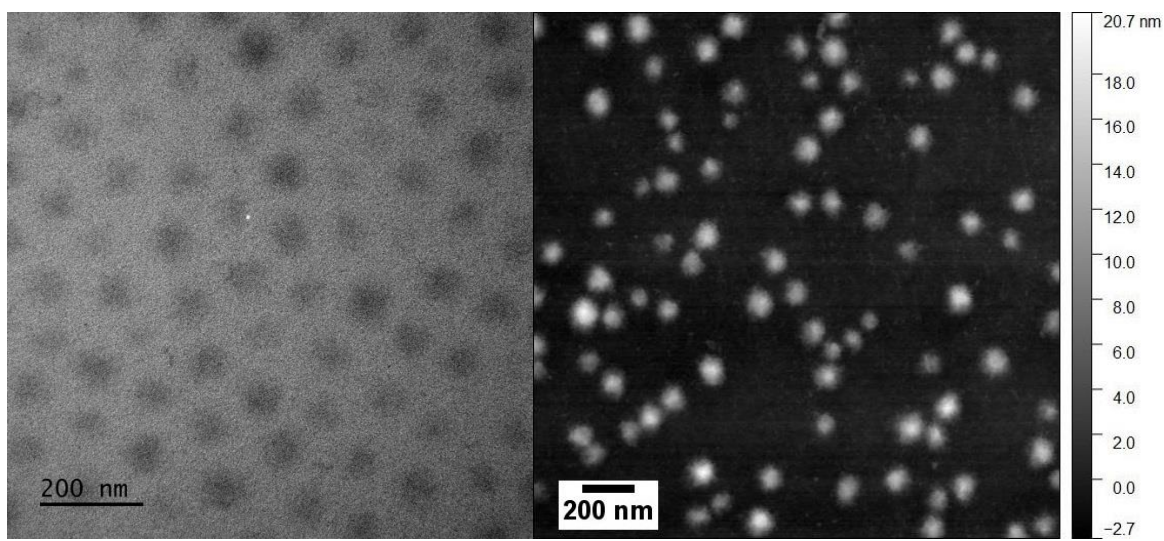

**Figure 2:** Real-space images of the microgel particles (CC = 10 mol%) in the dried state. Left: TEM, Right: AFM.



The size of the microgel particles as a function of temperature has been studied by PCS. The resulting hydrodynamic radius $R_H$ is shown in Figure 3. The volume phase transition temperature is found to be 44 °C, as expected for this monomer [41,67,68]. Several temperatures located in the approach of the transition temperature from below, as well as far below (at 15 °C) and above (at 55 °C) have been selected for the neutron scattering experiments. These temperatures are also indicated in Figure 3.

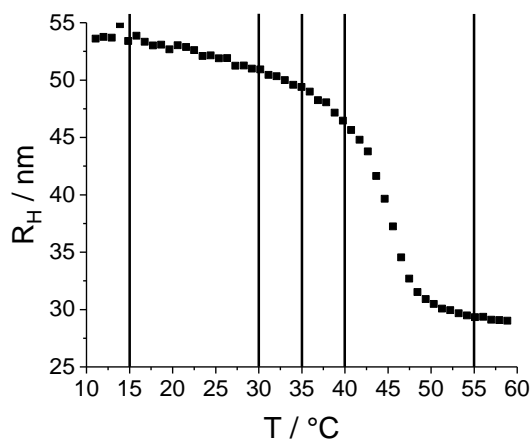

**Figure 3:** PCS measurements of the hydrodynamic radius $R_H$ of pNIPMAM microgel particles with 10 mol% CC as a function of temperature. The volume phase transition temperature is found to be 44 °C. The vertical lines indicate the temperatures at which the SANS measurements have been performed.

PCS allows the analysis of the total change in volume. As temperature increases, the hydrodynamic radius is found to decrease from $R_H = 54$ nm, first approximately linearly down to 47 nm at 40 °C. Then the microgel collapses to $R_H = 29$ nm, corresponding to a total volume decrease by a factor of about 6.5. The volume decrease depends on the CC and was studied elsewhere.[13] It may be concluded that the hydration of the microgel must decrease strongly, and this will be confirmed by our analysis below.

**A reverse Monte Carlo (RMC) algorithm for the determination of the form-free microgel density profile.** Modelling approaches of the microgel density profile $\Phi(r)$ in the literature discussed in the introduction suggest that a more general, form-free profile may might reveal additional features, and that such an approach might also be applied directly to more complex architectures, including polydisperse core-shell microgels currently studied in our laboratories. In this work, the radial monomer volume fraction profile of simple (i.e., a priori homogeneous) microgel particles is described by a discrete number of densities of monomers (fictitious, including crosslinker, see materials) in successive spherical shells of constant thickness. The volume fraction, number of monomers, contrast in $D_2O$, and monomer (repeat unit) volume are given for a standard sample (CC = 10%) in Table 1.



| | |
|---|---|
| **Microgel volume fraction** | $0.818 \cdot 10^{-3}$ |
| **Total number of fictitious monomers in microgel particle of average radius R** | $1.1 \cdot 10^5$ |
| **$\rho_{Mono}$ / $10^{10}$ cm$^{-2}$** | 1.0 |
| **$V_{Mono}$ / nm$^3$** | 0.195 |
| **Density / g cm$^{-3}$** | 1.10 |

**Table 1:** Monomer number and properties for a typical microgel particle (CC = 10 mol%, $R_H$ = 55 nm at T = 15 °C). The values of $\rho_{Mono}$, $V_{mono}$, and the density were obtained by the contrast variation shown in Figure 1. The average radius R corresponds to the calculation with shell thickness $\Delta R = 2$ nm.

The reverse Monte Carlo algorithm is illustrated in Figure 4a. Several different initial conditions have been tried out to check the robustness of our approach. It is shown in Fig. S2 of the SI that (very) different initial conditions hardly change the final profile, e.g. by distributing the monomers such that the average density is constant everywhere in the particle, or by completely filling the inner shells first. In practice, we have used outputs at high temperature as initial guess for the next lower one. Using Eqs. (2-4), the scattered intensity is calculated and compared to the measured one, via $\chi_i^2$:

$$\chi_i^2 = \frac{1}{N_q} \sum_1^{N_q} \left[ \frac{\left( I_{exp}(q_i) - I_{calc}(q_i) \right)}{I_{Err}(q_i)} \right]^2 \tag{5}$$

where $N_q$ represents the total number of q-values (typically 200), the sum over i extending over all data points of the experimental $I_{exp}(q_i)$ and theoretical $I_{calc}(q_i)$ intensities. $I_{Err}(q_i)$ is the error of the experimental intensity. The decision of acceptance takes into account two contributions, $\chi_i^2$ and $\chi_r^2$, where the latter describes the roughness of the density profile:

$$\chi_r^2 = \frac{1}{N_p - 1} \sum_1^{N_p - 1} [\Phi(j) - \Phi(j+1)]^2 \tag{6}$$

The sum extends over $N_p$, the total number of shells ($N_p$ = 30), and $\Phi$ is the monomer volume fraction in shell j. The acceptance is based on the sum of two contributions, with a weighting factor A which may favor either very good fits, or very smooth profiles [61]:

$$\chi^2 = \frac{(\chi_i^2 + A\chi_r^2)}{(1+A)} \tag{7}$$

A compromise ("stability plot") was sought by systematically varying A, cf. SI Figures S3 and S4. For our purposes, A = 100 gave stable and satisfying results both for the fits and in smoothness, and was used throughout this article. Note that this determination of A compensates for any variability in the definition of $\chi_i^2$, in case of different error bars on the intensity for example.

Then MC steps are performed, by moving groups of $N_{move}$ monomers randomly from one shell to another shell, following a set of rules: (a) The volume fraction of monomers in a cell cannot exceed 100 %, nor be negative. (b) The simulation starts with a rather coarse description, moving around groups of 100 monomers to start with, out of a total of typically $10^5$. This number is then progressively reduced to 1 during the simulation run in order to fine tune the profile, unless a satisfying $\chi_i^2$ is found. (c) The polymer density is only allowed to decrease with increasing radius. This avoids having empty shells in the middle of the particle. (d) The deviation between theoretical curve and measured intensity is expressed as $\chi^2$. MC moves are accepted if they decrease $\chi^2$, or if the increase is negligible with respect to $\chi_T^2$, expressed via a Boltzmann criterion $\exp(-\chi^2/\chi_T^2)$. $\chi_T^2$ plays the role of an effective temperature, which is set to a high value (five times the initial $\chi^2$, with a minimum of 1000), and which is then progressively decreased with the number of steps ("simulated annealing"). (e) The total number of MC steps is typically $10^5$-$10^6$. An example for a simulation run is shown in Figure 4b. The effective temperature $\chi_T^2$ is seen to decrease down to a lower bound, which is reached when the fit expressed by $\chi_i^2$ is judged to be acceptable (black squares), jointly with the roughness $\chi_r^2$ (green triangles). Then, the number of monomers per group is also frozen (blue circles in Figure 4b) and the profile and fit are averaged over the next 100 000 steps (Figure 4b). Note that RMC-parameters like the decrease of the number of monomers moved in each step affect only the speed of convergence, but not the final result (shown for comparison in Figure S8).

The outcome of this simulation procedure is a smoothly decaying monomeric volume fraction profile having a Fourier transform compatible with the observed scattering. Note that size polydispersity is included from the start. Only one profile is generated by the algorithm, corresponding to the nominal thickness of the shells. Size averages are then calculated as indicated in the methods section, i.e. stretching or compressing the shells, and the (unique) profile is modified until the averaged intensity is satisfyingly fitted.

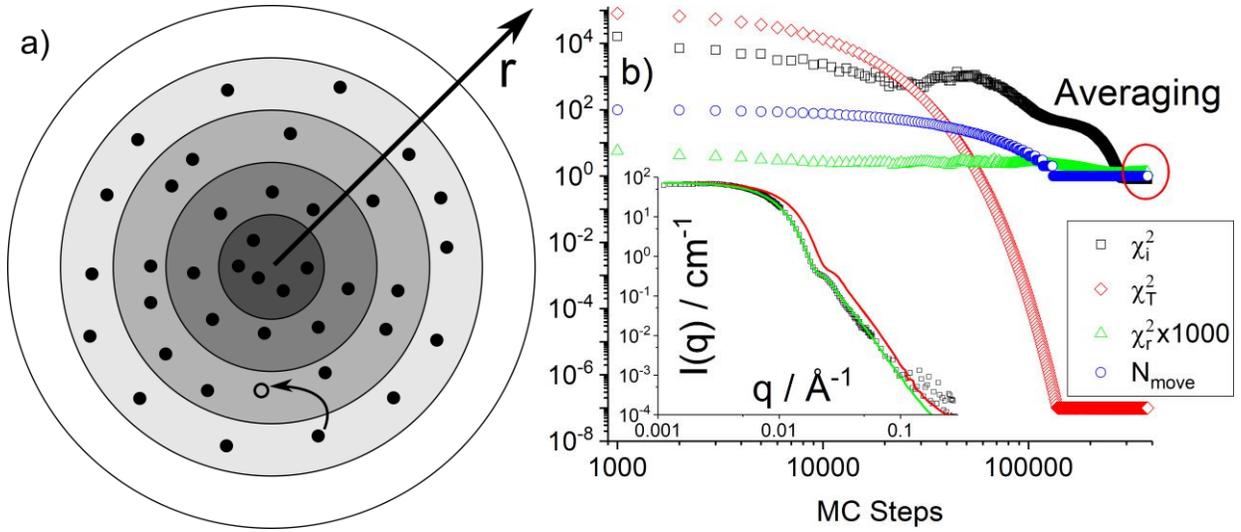

**Figure 4:** (**a**) Shell model of average microgel particle indicating moves of groups of monomers between shells. (**b**) Simulation parameters as a function of the number of MC steps ($\chi_i^2$, $\chi_T^2$, $N_{move}$, the number monomers moved in one step, as indicated in the legend. The lowest value is 1). Once $\chi^2$ falls below 1 the profiles and fits are averaged over the next 100 000 steps. In the inset, the initial guess for the intensity (in red) and the final simulation result (in green) are compared to the experimental intensity (black).

**Temperature-dependence of internal structure of microgels.** We have applied our form-free multi-shell model defined by Eq. (4) to pNIPMAM microgels containing CC = 10 mol% in D$_2$O, at different temperatures, covering the entire range of the volume phase transition temperature of this polymer (44 °C). The final profiles shown in this paper have been obtained by first describing the microgels at the highest temperature (55 °C), and using the resulting profile as initial condition for the next lower temperature, down to 15 °C. The profile for this last temperature was then recalculated again with different initial conditions, see SI, in order to check robustness.

The chain and profile contributions to a final fit of the experimental intensity are exemplarily shown separately in Figure 5a for T = 55 °C. We have chosen the highest temperature for illustration, as the particle surface is well-defined in this collapsed state. At intermediate q ($\approx 0.02$ Å$^{-1}$), this induces well-defined oscillations which are smoothed by polydispersity. The chain scattering is seen to be rather weak due to the limited particle hydration; it is responsible only for the high-q intensity. One might have worried about the oscillation being affected by the Guinier domain of chain scattering, but according to Figure 5a, this is clearly not the case. As size polydispersity (and the resolution function, see SI) is included, the description of this oscillation is thus entirely related to the shape of the density profile to be discussed below, and in particular to the steepness and curvature at the interface.

The entire data set as a function of temperature is shown in Figure 5b, and the corresponding fit parameters of the chains are given in Table 2. The total number of monomers has been fixed by adjusting the well-defined low-q intensity of the 55 °C-sample where the structure factor is negligible, and kept for all other T. It is given for microgels of nominal shell thickness ($\Delta R = 2$ nm) corresponding to the average radius in Eq. (1), and is higher (resp. lower) for bigger (resp. smaller) particles as described by the Gaussian size distribution. From the comparison of $I_0^{chain}$ to the low-q intensity of the microgel, it is deduced that approximately 1300 subchains make up the particle.

| Temperature / °C | Chain properties | | |
|---|---|---|---|
| | $I_0^C$ / cm$^{-1}$ | $R_g^C$ / nm | $\nu$ |
| 15 | 0.08 | 9.5 | 0.71 |
| 30 | 0.07 | 7.5 | 0.67 |
| 35 | 0.07 | 6.0 | 0.53 |
| 40 | 0.07 | 5.0 | 0.38 |
| 55 | 0.04 | 4.5 | 0.27 |

**Table 2:** Fitted chain parameters following Eqs. (2) for microgel temperature series (CC = 10 mol%).

$\nu$ was deduced directly from the slope of the chain contribution at high-q. It decreases from 0.71 at 15 °C to 0.27 at 55 °C. Thus the slope of the intensity increases from $q^{-1.4}$ at 15 °C to $q^{-3.7}$ at 55 °C, where the scattering approaches the one of a sphere with well-defined interfaces, and the corresponding Porod power law $q^{-4}$. The parameter $\nu$ evidences the evolution from rather extended chains of low fractal dimension at low temperature, to chains close to ideal Gaussian statistics at 35 °C, and a chain collapse at higher temperatures. As shown by the continuous decrease in the radius of gyration of chains $R_g^C$ which is our measure of the meshsize of the network, the network shrinks continuously with temperature, accompanying the changes in chain conformation. In parallel, the prefactor of the chain scattering given in Table 2 stays constant from low temperature up to the volume phase transition temperature of ca. 44 °C (see Fig. 3), and decreases only after, indicating that the amplitude of the network heterogeneity, or alternatively the mass of the subchains making up the network, remains conserved below 44 °C. Above this temperature, chains may start to stick together, and their visibility decreases.

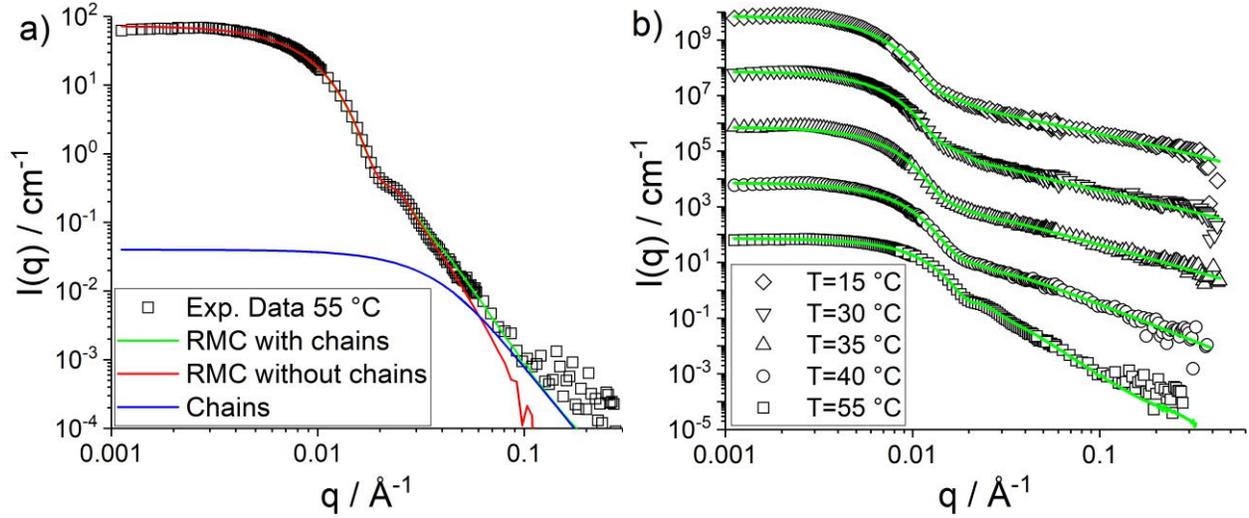

**Figure 5: (a)** Fitted chain and profile contributions to scattered intensity of a microgel (CC = 10 mol%) at 55 °C. **(b)** Temperature series from 15 to 55 °C of experimental intensities of the same microgel (symbols), compared to the model fits (solid lines).

The profiles $\Phi(r)$ corresponding to the parameters given in Table 3 and to the fits shown in Figure 5b have been plotted in Figure 6. They are represented as the monomer volume fraction in each of the concentric shells defining the geometry of the model. A first obvious validation is that all particles are entirely contained in these shells, i.e. the density goes to zero for sufficiently big radii, at all temperatures. In the swollen microgels at low temperature, the maximum spatial extension of the microgels is found, up to a radius of ca. 42 nm. This value can be compared to the hydrodynamic radius, $R_H = 54$ nm. The latter being higher indicates that the diffusion of the microgel particles is hindered by a low-density (i.e., hardly visible in SANS) corona of chains extending from the particle into the solution. It may be noted in Table 3 that the hydrodynamic radii are typically 20 nm above the radius of gyration $R_g$ of the particles as long as one stays below the transition temperature of 44 °C. Above, the $R_H$ is only 10 nm bigger than $R_g$, presumably due to the bad solvent condition of corona chains. This results in a collapse of the corona chains and thus a sharper surface, bringing $R_H$ closer to $R_g$, without however reaching the homogeneous sphere ratio of $R_g/R_H = \sqrt{3/5}$. The evolution of $R_g/R_H$ is shown in the inset in Figure 6. The curve has a similar shape and in particular a kink in the vicinity of the phase transition, as observed for pNIPAM by Deen et al.[42] Altogether, the evolution of the particle size with temperature found by SANS is compatible with the one found by PCS.

The size deduced from the profile can be compared to the radius of gyration of the particle. $R_g$ can be calculated from an integral over the density profile, or directly be determined from the Guinier region of I(q) as reported in Table 3, giving identical results within an error induced by the finite shell thickness. An additional way to quantify the spatial extension of the microgel particles is to take the radius $R_{50\%}$ where the volume fraction is half of its value in the center: $\Phi(R_{50\%}) = 0.5 * \Phi(0)$. As can be seen in Table 3, this $R_{50\%}$ is surprisingly close to $R_g$.

The higher spatial extension of microgels at low temperature is naturally accompanied by a lower monomer volume fraction in the center (i.e. at small r in Figure 6), due to conservation of the total number of monomers. The remaining volume being occupied by the solvent, the hydration levels can be directly read off from the low-r limits in Figure 6: Hydration = $1 - \Phi(r=0)$. Hydration is also reported in Table 3, and it can be seen to decrease from some 85 %v (in good agreement with recent findings by Dulle et al.[76] to about 42 %v with increasing temperature, accompanying microgel collapse. These values may be compared to those obtained on core-shell microgel particles[69], where the hydration was about 90 %v at low temperatures. One may also note that the evolution of the hydration values is compatible with the change in $R_g$, i.e. the model is self-consistent.

The radius of gyration $R_g$ in Table 3 is a global quantity describing the average mass distribution inside the microgel with a single parameter. The profile given by our form-free multishell analysis provides not only the water content, but also the shape and in particular steepness of the microgel-water interface. We have chosen to characterize the gradient of the interface by its width, which we have defined arbitrarily as the difference in radii $\Theta = R_{20\%} - R_{80\%}$, following the same definition as for $R_{50\%}$. The relative width, $\Theta/R_{50\%}$, is reported in Table 3. It is seen to evolve considerably, from highly extended, floppy microgels at low T, to smaller and more compact, well-defined microgel particles at high temperature. In absolute values, the width of the interface becomes even sharper, as not only the relative width decreases from typically 40 % to 15 %, but also the size decreases by some 30 %. In total, the absolute width decreases thus by roughly a factor of three. Moreover, the curvature of the profile at the surface of the particle is found to point upwards, and at low temperature, a small density gradient at the center of the particle is evidenced. This quantification of the evolution of the interface including thickness and curvature is one of the main findings of this article, which would have been difficult to obtain with predefined analytical profiles, let alone perfect spheres, as discussed in the last section of this article. The values for the monomer volume fraction and the interface thickness may be compared to the original studies by Stieger et al.[39]. These authors found typically above $\Phi = 40$ % in the center of the microgels at high temperature, and below 10 % at low T, depending on the crosslinking. These values compare favorably with our results. The relative width expressed by their Gaussian fuzziness evolves from very sharp at high temperature (some percent, a virtually box-like profile being suitable) to about 25 % at low T, depending again on crosslinking, i.e. the span is much larger than with our microgel particles. This may be due to different crosslink concentrations and a different overall size.

| Temperature / °C | $R_g$ / nm | $R_H$ / nm | $R_{50\%}$ / nm | Hydration / % | $\Theta/R_{50\%}$ |
|---|---|---|---|---|---|
| 15 | 34.5 | 53.4 | 29.0 | 84 | 0.38 |
| 30 | 32.0 | 50.9 | 29.0 | 77 | 0.35 |
| 35 | 31.2 | 49.4 | 27.0 | 75 | 0.48 |
| 40 | 26.9 | 46.5 | 24.0 | 66 | 0.42 |
| 55 | 19.9 | 29.3 | 21.0 | 42 | 0.14 |

**Table 3:** Particle parameters as a function of temperature. The particle radius of gyration $R_g$ was determined with a Guinier fit of the low-q region. The hydrodynamic radius $R_H$ is a result of the PCS measurements und $R_{50\%}$ was deduced from the density profiles in Figure 6. The hydration was calculated with $1 - \Phi(0)$ where $\Phi(0)$ was taken from Figure 6. $\Theta/R_{50\%}$ describes the steepness of the density profile in Figure 6 and was calculated as described in the text.

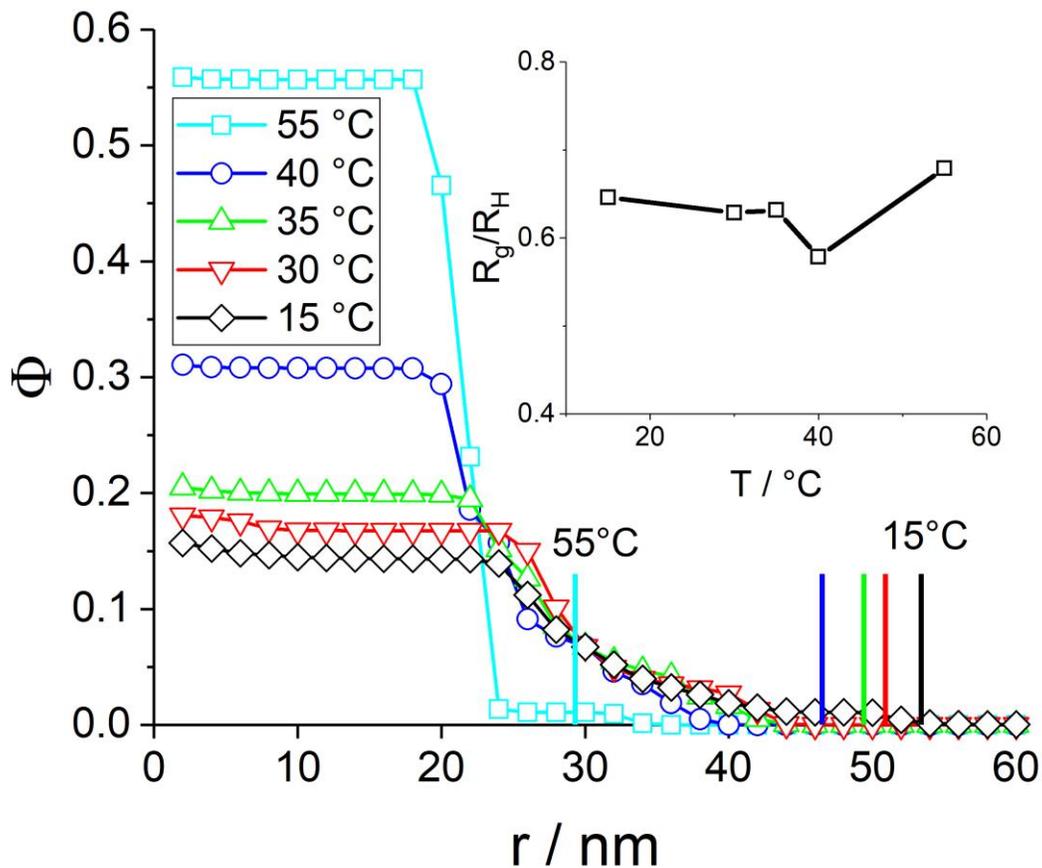

**Figure 6:** Density profile of the microgel (CC = 10 mol%) at different temperatures as indicated in the legend. The vertical colored bars indicate the hydrodynamic radii of the same samples. Inset: temperature-dependence of $R_g/R_H$.

**Effect of crosslinking on internal structure of microgels at high T.** Following the analysis of one system (CC = 10 mol%) at various temperatures, we have studied the influence of the crosslinker content in pNIPMAM microgels at 55 °C. The effect on the swelling behavior with T has been studied elsewhere by light scattering.[69] In Figure 7a the SANS intensities of a CC variation from 5 mol% to 15 mol% CC are compared to the reverse Monte Carlo fits, all of good quality. The form factor oscillation at ca. $q \approx 0.02$ $\text{Å}^{-1}$ is well pronounced for CC = 10 mol% and 15 mol%, and is weak for 5 mol% CC. The smearing of the oscillation at 5 mol% is due to a higher steepness inside the particle in spite of a better definition of the interface as deduced from the density profiles and summarized in Table 4. The density profiles obtained from the RMC simulation are shown in the SI. In Figure 7b, the particle sizes have been normalized to the one of the 10 mol% CC for comparison, because the microgel sizes vary for each synthesis and do not reflect differences in internal structure. The hydration in the center of the particle is found to increase from 2 % to about 50 % with increasing CC, contrarily to what one might have expected. At this high crosslinking values, this may be due to the higher chain flexibility of the less-crosslinked sample, allowing the chains to more efficiently occupy the available volume and expel the water molecules from the center of the particle. This observation is in agreement with the well-known decrease of the swelling capacity of pNIPAM microgels with increasing CC.[77] More crosslinks increase the mechanical constraints inside the network leading to an incomplete collapse and a structure still having space to accommodate water. As one can see in Figure 7b, a smooth gradient in hydration from the center towards the surface of the particles is found for 5 mol% CC, whereas at higher crosslinking density, there is no gradient in the center of the particles. The higher chain rigidity thus seems to confer a more homogeneous monomer density to the particles. The rigidity also seems to cause a difference in the evolution of the steepness of the profiles at the surface of the particle as given by $\Theta/R_{50\%}$ in Table 4. The most flexible, low-CC microgel particles have also the steepest gradient at the surface. All their chains are thus collapsed onto the particle, creating a well-defined polymer-water interface, whereas the more rigid chains still extend into the liquid, probably due to difficulties in folding back. It is noteworthy that the effect observed here is different from the independence of crosslinking found by Stieger et al.[39] at high temperature, whereas at low T (not studied here) microgels are swollen and higher crosslinking leads to denser particles. This difference is probably due to the higher crosslinking densities used in the present study, with its consequences on chain flexibility. As a final remark on this set of data, one may note that here the high generality of our form-free multishell approach allows to simultaneously describe a gradient inside the particles, and a different one at their surface, something which would have to be anticipated in any model of fixed shape function.

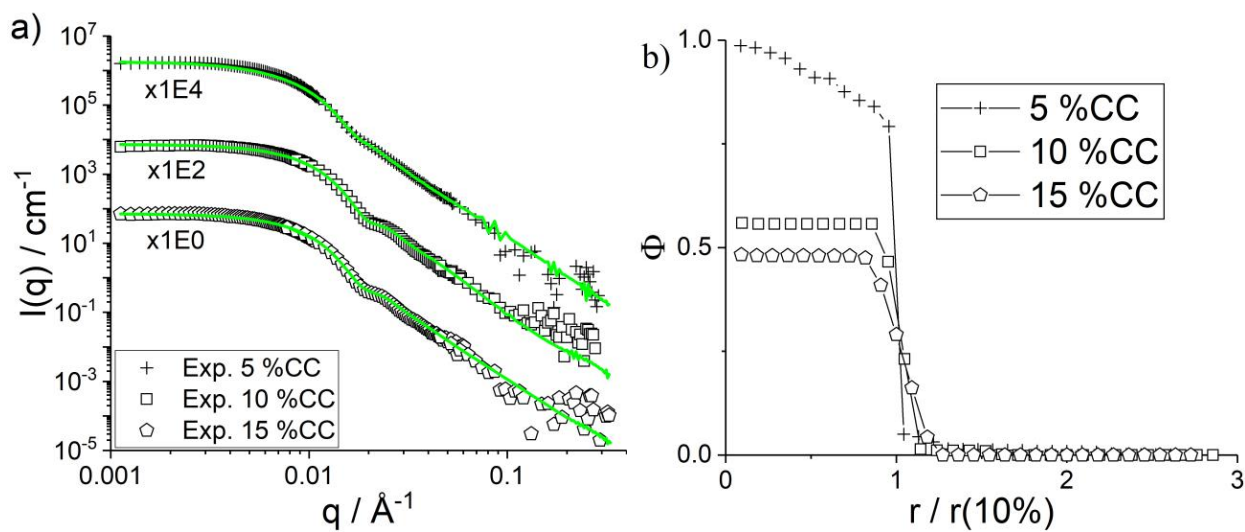

**Figure 7:** (**a**): SANS data with fits T = 55 °C shifted in intensity for clarity as indicated in the legend for three different crosslinker densities. The SLD of the fictitious monomer was adjusted (5 mol%: 0.96 $10^{10}$ cm$^{-2}$, 10 mol%: 1.00 $10^{10}$ cm$^{-2}$, 15 mol%: 1.03 $10^{10}$ cm$^{-2}$). (**b**) pNIPMAM monomer volume fraction profiles for the same samples, renormalized to the radius of the 10% CC sample for comparison.

| CC / mol% | $R_g$ / nm | $R_H$ / nm | Hydration / % | $\Theta/R_{50\%}$ |
|-----------|-----------|-----------|---------------|-------------------|
| 5 | 23.9 | 34.3 | 2 | 0.09 |
| 10 | 19.9 | 29.3 | 43 | 0.14 |
| 15 | 21.9 | 32.2 | 52 | 0.17 |

**Table 4:** Particle parameters for a CC variation. The particle radius of gyration $R_g$ was determined with a Guinier fit of the low-q region. The hydrodynamic radius $R_H$ is a result of the PCS measurements und $R_{50\%}$ was deduced from the density profiles in Figure 6. The hydration was calculated with $1 - \Phi(0)$ where $\Phi(0)$ was taken from Figure 7b. $\Theta/R_{50\%}$ describes the steepness of the density profile in Figure 7b and was calculated as described above.

**Form-free analysis of deuterated microgels.** As a preparation for upcoming experiments with deuterated core-shell microgels analyzed with the form-free multishell model, we have performed an identical microgel synthesis using two different deuterated monomers, called D7 and D12 reflecting their degree of H-D-substitution, and hydrogenated crosslinker. The SANS experiments have been performed in H$_2$O, and scattering length densities determined by contrast variation to 4.45·$10^{10}$ cm$^{-2}$ and 7.62·$10^{10}$ cm$^{-2}$ for D7 and D12, respectively. The scattered intensities are shown in Figure 8a, and the resulting monomer density profiles in Figure 8b. The main result is that hydration levels are much higher, even at high temperatures, indicating that the polymerization and crosslinking reaction have followed a different kinetics as compared to the hydrogenated monomers. It can be expected that the deuterated monomers react slowlier compared to their hydrogenated counterparts. This should lead to an even steeper crosslinker gradient compared to the "normal" microgels. At low temperature, the shape of the profiles is found to be more extended into the solution, microgel particles are fuzzier. This analysis of pure D-microgels thus gives valuable information on how to proceed with microgel particles containing both H- and D-monomers.

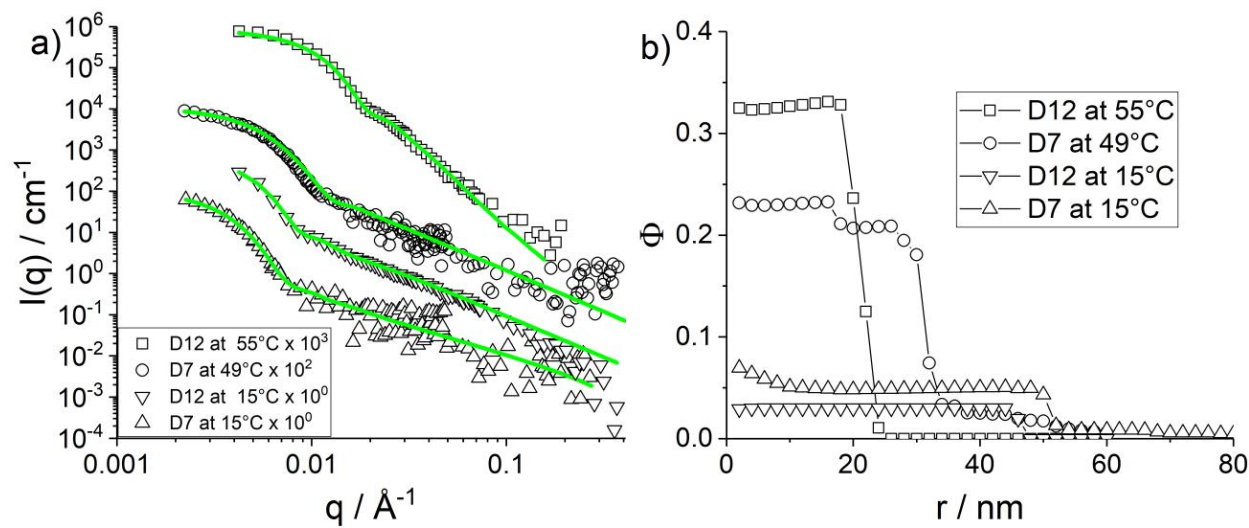

**Figure 8: (a)** Intensity I(q) vs. q of D7 (resp. D12) microgels with 10 mol% CC at 15 and 55 °C, compared to RMC fits. The volume fraction Φ of the D12 samples was 0.00818 and for the D7 samples 0.000818 **(b)** Corresponding density profiles.

**Comparison of the RMC-multishell model to fuzzy-sphere-like models.**

In the following, we compare our results obtained with the general form-free multi-shell algorithm to prominent models from literature [39,51,78]. As already mentioned in the introduction, a similar form-free algorithm has been proposed [62], but not applied to spherically symmetric systems, and in particular not to microgels.[61,63] For the sake of comparison, all model predictions have been calculated using the same relative particle polydispersity (see Eq. (1) and SI) of 20.4 %. The simplest model is the homogeneous sphere model[78], i.e. the intensity of which corresponds to the form factor of a sphere, which by definition does not describe any gradient, in particular at the interface. This simple model has been shown to work rather well for microgels in the collapsed state (here 55 °C). At high q, its characteristic Porod law (q$^{-4}$) describes well-defined surfaces. Its corresponding fit shown on the left, Figure 9a, (including a weak polymer contribution as described by Eqs. (2), ν = 0.27, indicating poor solvent) is surprisingly good, as already recognized by Stieger et al.[39]. The zoom shown in the inset, however, demonstrates that the sphere model with polydispersity still does not perfectly fit the oscillation. The same is true with the Boon-Schurtenberger model, which underestimates the intensity around the oscillation shown in the inset [51]. Its Flory-Rehner approach is a priori not appropriate for high-T collapsed states, because it focuses on the visibility of solvated polymer chains. The model still works correctly, presumably because of the non negligible hydration in the core, as shown in Figure 9b. The Gaussian fuzzy-sphere model presents a good fit, slightly overestimating the intensity at the oscillation, while the fit of our multi-shell and of the piecewise-parabolic models are seen to be even better. In Figure 9b, the corresponding volume fraction profiles are shown. All volume fractions in the core of the particles reach a similar value, around 60 %, and the decrease of the profile occurs in the same size range. The shapes of the profiles, however, differ considerably. The sphere-model is easily recognized by its rectangular shape in the radial representation. The density profile of the Boon-Schurtenberger model is also plotted, with its characteristic progressive decrease described by a third-degree polynomial. The results of two best fits, the parabolic fuzzy-sphere model and the multi-shell model, can be directly compared in Figure 9b. Almost identical core hydration values are found. The width of the profile is seen to be non-zero, although it is still rather steep, and inside the particle a homogeneous, constant profile is reached by both the multi-shell and the fuzzy-sphere models.

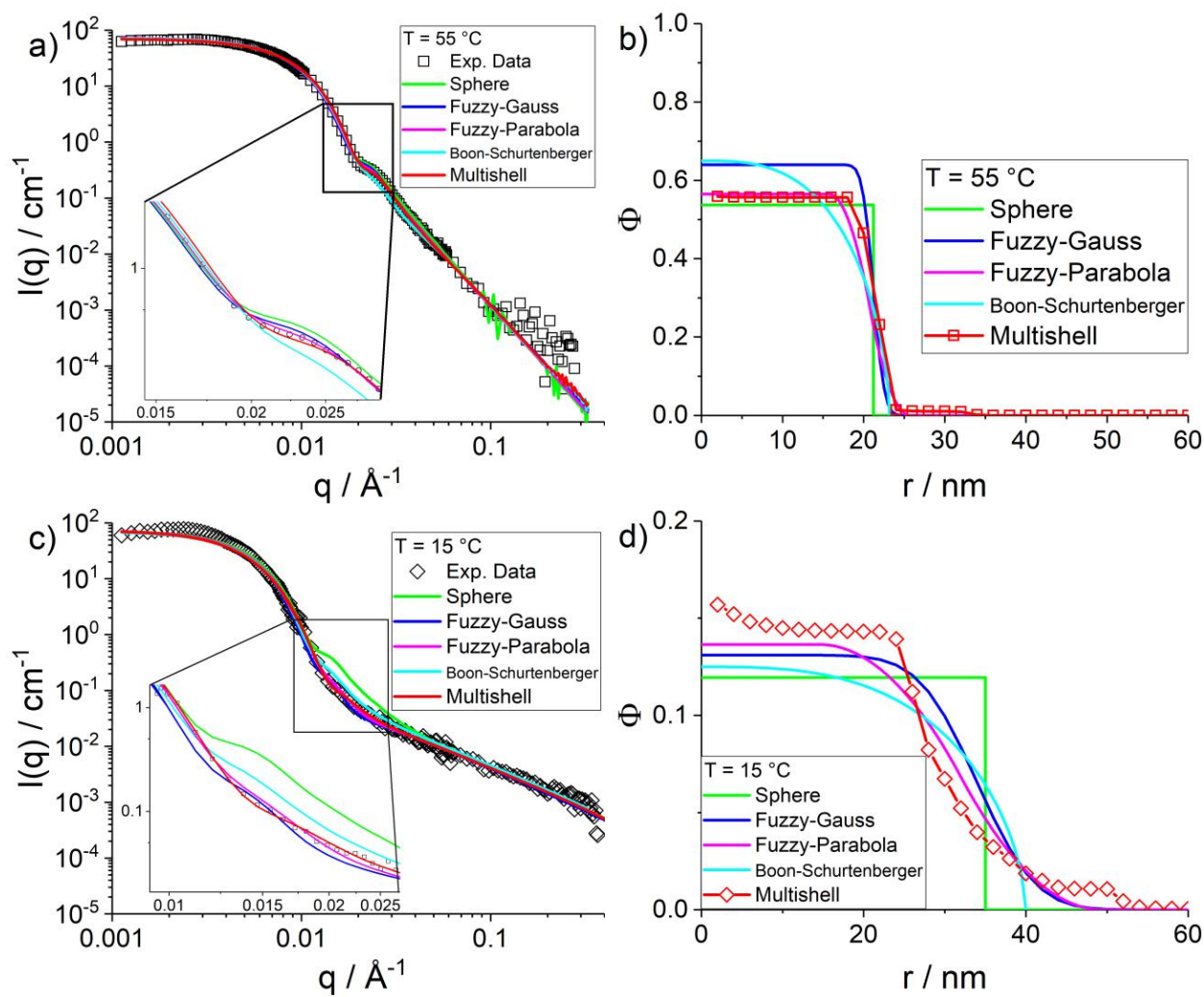

**Figure 9:** Comparison of experimental intensities (10 mol% CC) with other models as indicated the legend. **Left**: I(q) vs. q, **right**: density profiles, **top**: 55 °C, **bottom**: 15 °C.

For the lowest temperature of 15 °C, model intensities are compared in Figure 9c to the measured one. The sphere model presents the same oscillation as at 55 °C, as steepness of the profile cannot be adapted. Thus the sphere model does describe our data at 55 °C approximatively, but completely fails in the swollen state of the microgels, as anticipated and already seen by others.[39,51] The Boon-Schurtenberger model is again slightly off in the oscillation region, this time overestimating the intensity. The multi-shell and the fuzzy-sphere models are again quite successful, but one may note in the zoom that some oscillations not found in the experiment persist in the Gaussian fuzzy-sphere fit, while the other two models perform equivalently. On the scale of Figure 9d, the monomer volume fraction profiles show some differences, although all models predict comparable hydration levels in the center, corresponding to monomer volume fractions between 12 % and 16 %, instead of 60 % at 55 °C. The steepness of the profile at the surface of the particles on the other hand is much lower than at 55 °C. The profiles of both the fuzzy-sphere models and the multi-shell model in Figure 9d have a similar spatial extension, but the higher flexibility of our model results in a non-constant profile inside the particle, which might be reminiscent of a radially non homogeneous crosslinking density and was also observed by dSTORM[52]. Moreover, contrarily to the 55 °C data, the curvature of the fuzzy- and multishell-profile at the surface is opposite, which again would have had to be anticipated to be described by a fixed-form model. Together with the possibility to describe the internal gradient and the particle surface independently and freely, this represents the main outcome of this article. This flexibility, precise description, and the above-mentioned compatibility to deuteration will be crucial for further studies of core-shell microgels where one part is hydrogenated and the other deuterated.

## Summary and conclusion

We have applied a form-free description of the monomer density profile of spherically-symmetric, polydisperse microgel particles to SANS intensity curves, together with a polymer scattering term which is only of importance at high q, and the exact form of which does not influence our results on the profile. The analysis of this term shows that polymer chains are not only visible up to the volume phase transition temperature, but also above, accompanied by a decrease in visibility and an increase in fractal dimension, due to the poor-solvent conditions. Our description is coarse-grained, and the nominal thickness of the shells is 2 nm. We have implemented reverse Monte Carlo steps of monomers moving from one shell to the other, progressively optimizing both the agreement with the scattered intensity, and the smoothness of the profile. The results have been shown to be robust with respect to various initial conditions. The polydispersity and resolution effect have been taken into account based on independent AFM pictures, and the characteristics of the SANS beamlines, respectively. Working with known polydispersity increases the trustworthyness of the resulting density gradients. One may note that the presence of interactions in particular in the swollen state sometimes leads to very weak structure factors in the low-q range. As the low-q intensity is fixed by the number of monomers in each particle, as well as the scattering length density as determined independently by contrast variation, this does not have any consequences for our approach.

Our method has been applied to a microgel series in temperature, from 15 °C (swollen) to 55 °C (collapsed), the transition temperature of pNIPMAM being 44 °C as found by dynamic light scattering. All fits are of very high quality, and turn out to be better than most existing models, in particular around the first form factor oscillation, with the exception of the piece-wise parabolic fuzzy-sphere model. However, our method is more general, which is decisive for the description of gradients inside the particles. The resulting profiles not only provide the internal hydration of the microgels, but also a measure of the gradient at the interface (steepness), as well as the curvature of the profile, which was found to be positive. Moreover, in the highly swollen state, the density inside the microgel particles was found to be non-homogeneous, with an increase in density towards the center of the particles. This may be a possible signature of a heterogeneous crosslinking reaction within the particle.[35] The effect of the total crosslinking density was also studied, between 5 mol% and 15 mol%, and surprisingly the less crosslinked samples displayed the lowest hydration values, as well as the steepest gradient at the particle surface, presumably due to the possibility of the molecules to backfold more efficiently.

Our model is highly general in the sense that it can be applied to microgels of arbitrary composition, and in particular microgels containing both H- and D-monomers. We are currently working on H-D core-shell microgels, the core of which is exactly the one studied in this article. We will thus be able to quantify the extent of compression of the core by the shell, using our approach. In order to prepare this, we have included in the present article a study of the effect of deuteration, comparing H, D7 and D12 monomers. Preliminary experiments with deuterated microgels hint at differences in chemical reactivity, leading to more floppy and more hydrated microgels than with H-monomers. It is also promising to use our method to look into the monomer profiles as a function of particle size, where non-spherical structures[10] have been reported by some of us in the past, and which still need to be understood[79]. Finally, it is hoped that our method will also be compared to density measurements obtained by high-resolution microscopy of much bigger microgel particles [52].

**Acknowledgements:** The authors are thankful for support by the joint ANR/DFG CoreShellGel project, Grant ANR-14-CE35-0008-01 of the French Agence Nationale de la Recherche, and Grant HE2995/5-1 by Deutsche Forschungsgemeinschaft. This work is based upon experiments performed at the KWS-1 instrument operated by JCNS at the Heinz Maier-Leibnitz Zentrum (MLZ, Garching, Germany) and PA20 at LLB (CEA, Saclay). We thank Matthieu George (L2C Montpellier) for the help with AFM measurements, and Ina Ehring (Bielefeld) for the help with the syntheses, and Oliver Wrede for help with the design of the journal cover.

**Supporting Information:** Polydispersity, fits and their stability, chain contribution to scattering, additional density profiles, details on contrast variation and on light scattering.

**For Table of Contents Only:**

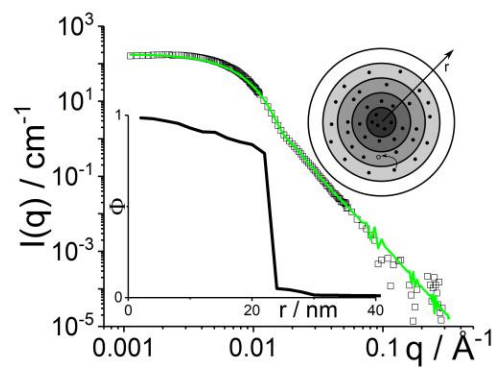